\documentclass[twocolumn,a4paper,10pt]{article} 
\usepackage[french]{babel}
\usepackage[T1]{fontenc}
\usepackage{epsfig}
\begin{document}
\pagestyle{plain}
%
%
\title{
{\Large \bf 
Rôles des contraintes sur les signaux de transition de phase.
}
}
 
\author{
{\bf 
Vincent Régnard <regnard@ganil.fr>
} \\ 
G.A.N.I.L., Bd. H. Becquerel B.P. 5027 F-14021 Caen cédex, France \\
L.P.C. , Bd. Mal Juin F-14050 Caen cédex, France 
}

\date{}

\maketitle

%
\begin{center}
{\bf 
Ce papier présente la notion de signal de transition de phase pour un système de taille finie. Il se concentre sur le rôle des contraintes physiques sur ces signaux et la robustesse quant au changement d'ensemble statistique. Des résultats obtenus avec un modèle de gaz sur réseau sont présentés en analogie avec les observations effectuées en collisions d'ions lourds aux énergies intermédiaires.
} 
\end{center}
%
\section{Introduction}
%
Les transitions de phase dans les systèmes finis ont été étudiées dans de nombreux cadres théoriques. Des approches simples comme le modèle de gaz sur réseau permettent d'étudier certaines propriétés comme la coexistence de phases ou la criticité et décrire de façon qualitative les processus physiques comme la multifragmentation des noyaux atomiques ou des agrégats d'atomes.
 La description de la transition de phase dans ces systèmes diffère de la phénoménologie classique que nous connaissons bien. Pour ces systèmes de taille finie, les fluctuations des observables sont importantes, les fonctions de partition contenant un nombre fini de termes ne permettent pas aux potentiels thermodynamiques de diverger, le rôle des interfaces n'est plus négligeable et surtout les ensembles statistiques ne sont pas équivalents. Une nouvelle description des transitions de phase pour ces sytèmes doit être formulée.
Après une présentation du modèle de gaz sur réseau et quelques rappels de physique statistique, je présenterai les effets des contraintes physiques appliquées au système, c'est-a-dire le choix d'un ensemble statistique, sur deux signaux de transition de phase.%
%
\section{Le modèle Lattice-gas}
%
%
Le modèle du gaz sur réseau est une extension du modèle d'Ising.
Ce modèle suppose un réseau multidimensionnel (3 dimensions dans notre cas) comprenant N sites caractérisés par un nombre d'occupation $n_{i}$ (absence (0) ou présence (1) d'une particule sur le site $i$) ainsi qu'une impulsion $\overrightarrow{p_{i}}$. La donnée des N nombres d'occupation et impulsions associées définit une configuration microscopique $\{n\}$. 
Le Hamiltonien que nous utilisons dans nos simulations s'écrit de la façon suivante:
$$
H(\{n\})=-\frac{\epsilon}{2}\sum_{i \ne j}^{N}n_{i}n_{j}+\sum_{k=1}^{N} n_{i}\overrightarrow{p_{i}}^{2}
$$ 
La première somme traduit une interaction attractive constante entre plus proches voisins sur le réseau (6 au maximum en 3 dimensions). 
Le paramètre $\epsilon$ peut être vu comme la profondeur du puits de potentiel d'une interaction générique à courte portée. 
Cette première somme est responsable de la cohésion du système et joue un rôle de première importance. La deuxième somme prend en compte l'agitation du système à travers l'impulsion des particules du réseau donnant ainsi des degrés de liberté supplémentaires au système. Cette agitation contribue à la dissociation du système en fragments. 
Les fragments sont identifiés à l'intérieur de chaque configuration microscopique par un algorithme de reconnaissance du type Coniglio-Klein~\cite{coniglio}. L'observation du système consiste ensuite à évaluer des observables pertinentes calculées sur un état microscopique du système.
%
%
\begin{table*}[bth]
\begin{center}
\begin{tabular}{||c|*{2}{c|}|}
\hline
Distribution statistique & Grand Canonique & Canonique  \\
\hline
\hline
Multiplicateurs de Lagrange & $\mu,T$ & $T$ \\
\hline
Lois de conservation & $<N>,<E>$ & $N,<E>$  \\
\hline
Distribution de probabilité & $P_{n}=\frac{1}{Z_{\mu,\beta}} e^{-\beta(H_{n}-\mu N_{n})}$ &  $P_{n}=\frac{1}{Z_{\beta}} e^{-\beta H_{n}}$ \\
\hline 
\end{tabular}
\caption{Tableau récapitulant les propriétés des ensembles statistiques utilisés.}
\label{tab}
\end{center}
\end{table*}
\section{Ensembles statistiques}
%
Les contraintes physiques extérieures sur le système étudié sont appliquées par le biais de multiplicateurs de Lagrange. L'observable $A_{l}$ (variable extensive) associée à $\lambda_{l}$ (paramètre intensif) se retrouve alors distribuée dans un intervalle. Par exemple le multiplicateur de Lagrange associé au volume est la pression, celui associé au nombre de particules le potentiel chimique ou encore celui associé à l'énergie la température. Pour toute observable nous pouvons définir un paramètre de Lagrange (conjugué canonique) qui permettra d'obtenir la distribution de cette observable à l'équilibre de Gibbs, c'est à dire celle maximisant l'entropie statistique sous contraintes. Cette méthode permet de traiter de façon équiprobable l'information dont on ne dispose pas, le seul biais introduit étant précisément l'information pertinente.\\  
Un algorithme Monte-Carlo de type Metropolis~\cite{metropolis} permet d'échantilloner différents ensembles statistiques afin d'obtenir les distributions statistiques de microétats satisfaisant les contraintes.
Il est possible d'ajouter à ces contraintes des lois de conservations (distribution de Dirac de l'observable conservée).
Nous considérerons les ensembles grand-canonique dans lequel le potentiel chimique et la température sont fixés (ie. nombre de particules et energie définis en valeur moyenne) ainsi que l'ensemble canonique où la masse est fixée et l'énergie est définie en valeur moyenne.
Les distributions de probabilité pour chaque ensemble statistique sont rappelées dans le tableau \ref{tab}.
%
%
\section{Bimodalité}
%
Une fois les distributions statistiques de microétats obtenues, les projections de la densité d'état sur des directions d'observation jugées pertinentes permet d'étudier la physique associée au système.

F. Gulminelli et P. Chomaz ont proposé~\cite{cho99} d'utiliser la topologie de la distribution des évènements dans l'espace d'observation pour signer la transition de phase pour les systèmes de taille finie. Dans la zone de coéxistence de phase une bimodalité (deux maxima) dans la distribution de probabilité d'une observable permet d'identifier sans ambigüité chacune des phases; cette observable joue alors le rôle d'un paramètres d'ordre.
 La projection de la densité de probabilité dans des espaces d'observation pertinents permet d'identifier l'appartenance des évènements à chacune de ces phases.

 La densité de matière est le paramètre d'ordre naturel de la transition de phase liquide-gaz à la limite thermodynamique. Il est toutefois possible de trouver d'autres paramètres d'ordre corrélés à la densité; en particulier on peut s'attendre à ce que la taille du plus gros des fragments ($A_{big}$) puisse servir de paramètre d'ordre. 

\begin{figure}[tb]
\centering\epsfig{file=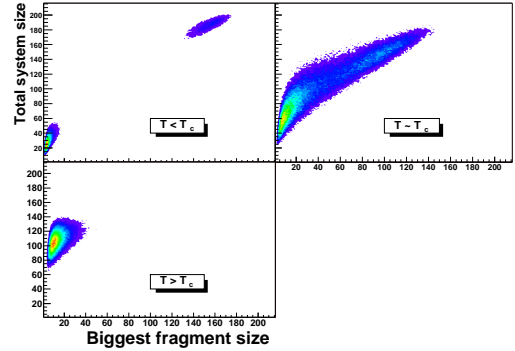,width=7cm}
\caption{Densité de probabilité grand-canonique projetée dans l'espace {\it``nombre total de fragments''} (ordonnées) versus {\it``taille du plus gros fragment''} (abscisse) pour différentes températures (dans un réseau 6x6x6)
}
\label{density} 
\end{figure}

 La figure (\ref{density}) représente la distribution des évènements grand-canoniques (à la valeur critique du potentiel chimique $\mu = 3\epsilon$) dans l'espace {\it``taille totale du système (ie. densité) versus taille du plus gros fragment''} et montre la corrélation entre ces deux observables.
Cette figure montre que même en système de toute petite taille la coéxistence de phase est reconnaissable par la bimodalité de la distribution du paramètre d'ordre. La claire corrélation entre $A_{big}$ et et la masse totale du système $A_{tot}$ démontre en plus que $A_{big}$ peut être utilisé comme paramètre d'ordre.
Cette observable est facile à manipuler dans les simulations numériques et accessible assez facilement expérimentalement. La densité étant beaucoup plus difficile (voire impossible du fait de la difficulté d'accéder au volume) à mesurer, l'étude de cette observable est d'un grand interêt pour faire des analogies ou comparer les résultats des modèles avec l'expérience.
A basse température la distribution de ces observables est bimodale (la projection sur chaque axe révèle deux maxima). Les évènements denses (grande masse totale) et ayant un gros plus gros fragment $A_{big}$ sont identifiés comme appartenant à la phase de type liquide tandis que le deuxième lot d'évènements (faible densité et petite taille du plus gros fragment) correspond à la phase gazeuse. En élevant progressivement la température du système, ces deux lots d'évènements se rapprochent l'un de l'autre pour fusionner à la température critique. Enfin pour les températures surcritiques la distribution devient monomodale (un maximum), une seule phase fluide subsiste, avec un plus gros fragment de petite taille.
Ainsi le passage d'une distribution bimodale à une distribution monomodale signe le passage par un point critique correspondant à la transition de phase du second ordre.
\begin{figure}[tb]
\centering\epsfig{file=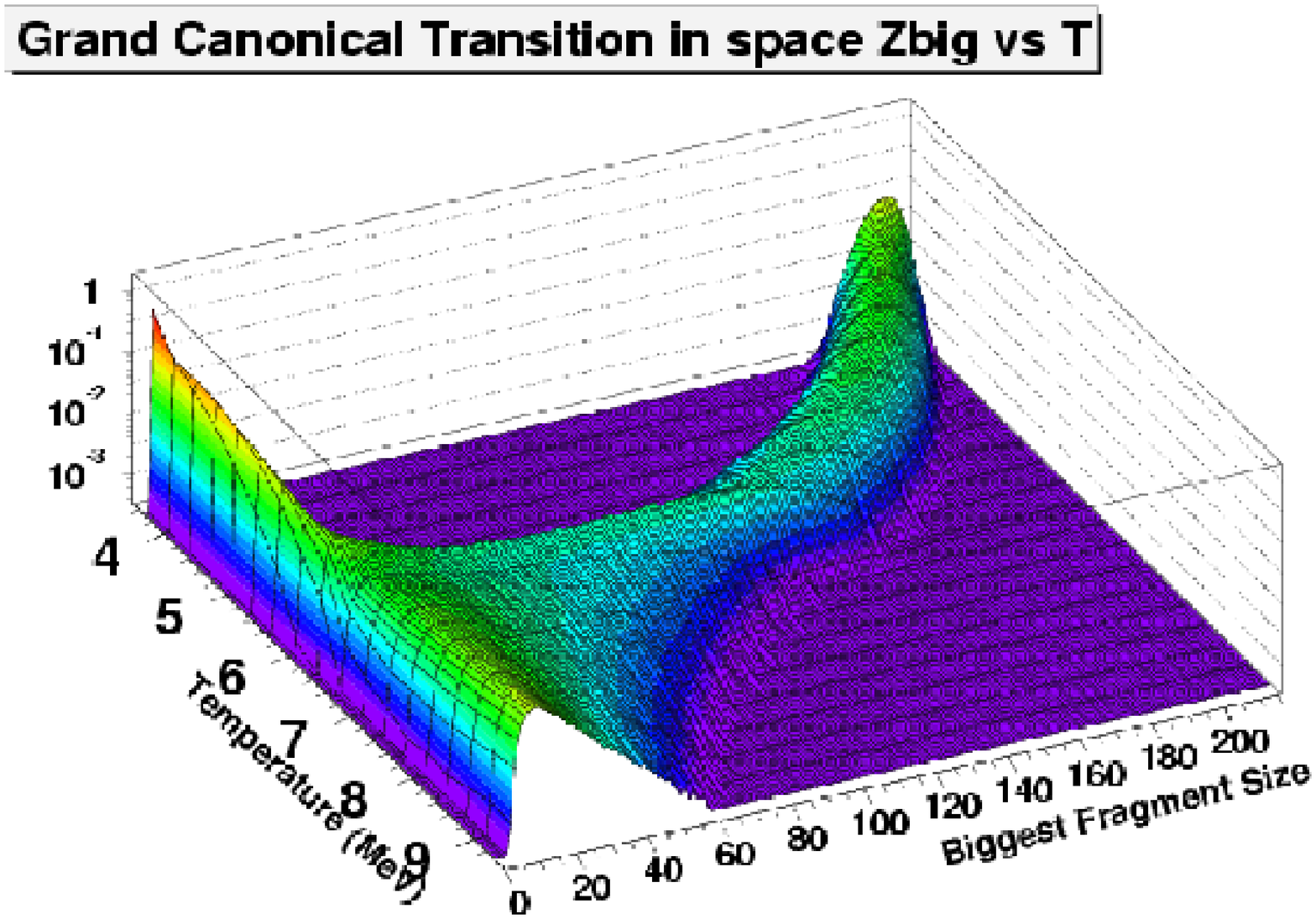,width=8.0cm}\\
\centering\epsfig{file=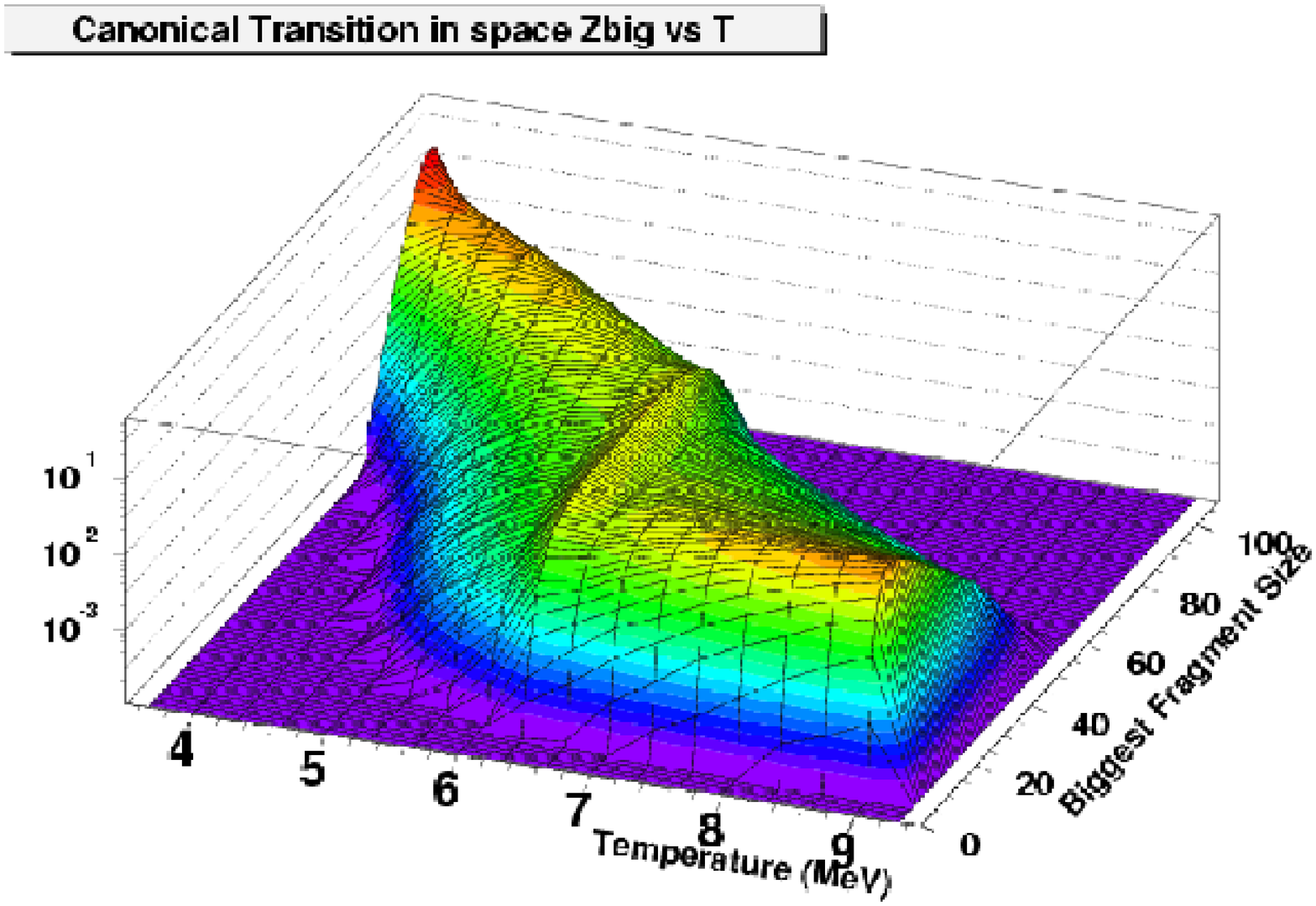,width=8.0cm}
\caption{Distribution de taille du plus gros fragment pour différentes températures dans l'ensemble grand-canonique (haut) et dans l'ensemble canonique (bas). 
}
\label{3D} 
\end{figure}

La figure (\ref{3D}) (graphique supérieure) schématise la bifurcation caractéristique d'une transition de phase. Cette figure montre la distribution grand-canonique du plus gros fragment pour différentes températures. Le passage de la distribution bimodale de ce paramètre d'ordre à une distribution monomodale est bien localisée et permet d'identifier le point critique. 
Le graphique inférieur sur cette même figure présente les distributions du plus gros fragment dans la même gamme de températures pour l'ensemble canonique. Dans cet ensemble la masse a été fixée, c'est-à-dire le nombre de sites occupés est maintenu constant, alors qu'il était libre de fluctuer (sous la contrainte du potentiel chimique) dans l'ensemble grand-canonique. Si l'existence d'une loi de conservation empêche évidement l'utilisation de la masse totale comme paramètre d'ordre, la bimodalité de la distribution de taille du plus gros fragment a aussi disparu dans cet ensemble. La contrainte sur l'observable {\it ``masse totale du système''} a fondamentalement changé la topologie de la distribution de $A_{big}$ à cause de la forte corrélation entre $A_{big}$ et $A_{tot}$. Le plus gros fragment n'est plus un paramètre d'ordre pour cet ensemble et ne permet plus de caractériser la coéxistence de phase.         
%
\section{Invariance par changement d'échèlle}
%
Il a été proposé de nombreux autres moyens de mettre en évidence une transition de phase. R. Botet et M. Ploszajczak~\cite{bot00} ont proposé par exemple d'utiliser la théorie des fluctuations universelles ou $\Delta-scaling$. Cette théorie démontre que dans un phénomène critique la distribution du paramètre d'ordre est invariante par changement d'échèlle du système. 
Cette théorie a des applications en physique nucléaire pour ce qui concerne la phénoménologie de la multifragmentation des noyaux excités à des énergies comparables à leur énergie de liaison, si la taille du plus gros fragment est paramètre d'ordre de la transition de phase associée à la multifragmentation et si celle-ci est interprétable comme un phénomène critique (ie. a lieu en proximité du point critique du diagramme de phase).
 On attend dans ce cas que la forme de la distribution du plus gros fragment (moyenant une transformation d'échèlle) reste la même lorsque la taille du système est modifiée traduisant le caractère universel de cette distribution. 
Le passage par un point critique se traduit par la vérification du $\Delta-scaling$ de la distribution normalisée $P_{<m>}(m)$ du paramètre d'ordre $m$ pour différentes tailles de systèmes :
\begin{equation}
<m>^{\Delta} P_{<m>}(m)= \Phi (z_{\Delta}),  
z_{\Delta} = \frac{m - m^{*}}{<m>^{\Delta}}
\label{botet}
\end{equation}
où $m^{*}$ et $<m>$ représentent la valeur la plus probable et la valeur moyenne du paramètre d'ordre et $\Phi$ la fonction d'échèlle.
Dans une phase désordonnée les distributions des paramètres d'ordre doivent vérifier $\Delta=1$, le passage par un point critique est signé par $\Delta=\frac{1}{2}$.
J.M. Carmona {\it et al.} ont montré~\cite{car02} que cette théorie est vérifiée dans le modèle de gaz sur réseau.
Cette technique est cependant difficle à mettre en oeuvre car la qualité de la loi d'échèlle (\ref{botet}) dépend évidemment de la valeur de $\Delta$ qui peut à priori varier entre 0 et 1 et n'est pas connue avant l'analyse.
 
En ce qui concerne les données expérimentales, le nombre limité d'espèces nucléaires rend impossible une application directe de la méthode. Les propriétés d'invariance d'échèlle ont alors été étudiées pour une même collision d'ions lourds en variant la violence de la collision~\cite{bot01} et un changement de régime a été observé de  $\Delta=1$ vers $\Delta=\frac{1}{2}$.
L'ensemble statistique (tri expérimental) dans lequel est effectué cette analyse s'apparente à un ensemble gaussien intermédiaire entre un ensemble microcanonique et un ensemble canonique.

En notant $P(A_{big})$ la distribution du plus gros fragment, $<A_{big}>$ et $\sigma_{A_{big}}$ la moyenne statistique et la variance de cette distribution, l'équation (\ref{botet}) peut être écrite de façon équivalente par l'ensemble des deux relations
\begin{equation}
\sigma_{A_{big}} P(A_{big}) = \Psi (z_{PMS}) = \Psi (\frac{A_{big}-<A_{big}>}{\sigma_{A_{big}}})
\label{pms1}
\end{equation}
\begin{equation}
\frac{<A_{big}>}{\sigma^{2\Delta}_{A_{big}}} = Constante
\label{pms2}
\end{equation}
 L'équation (\ref{pms1}) exprime la transformation qui permet d'obtenir les distributions centrées $\Psi(z_{PMS})$ de la variable réduite $z_{PMS}$. Ceci constitue une première étape qui ne suppose pas la valeur de $\Delta$.
L'équation (\ref{pms2}) exprime une relation affine entre les logarithmes des valeurs moyennes et des variances des distributions de plus gros fragments
$$
ln(A_{big}) = Cte - \Delta ln(\sigma_{abig})
$$
qui permet, si la loi d'échèlle (\ref{pms1}) est vérifiée, d'établir la valeur de $\Delta$.
\begin{figure}[tb]
\centering\epsfig{file=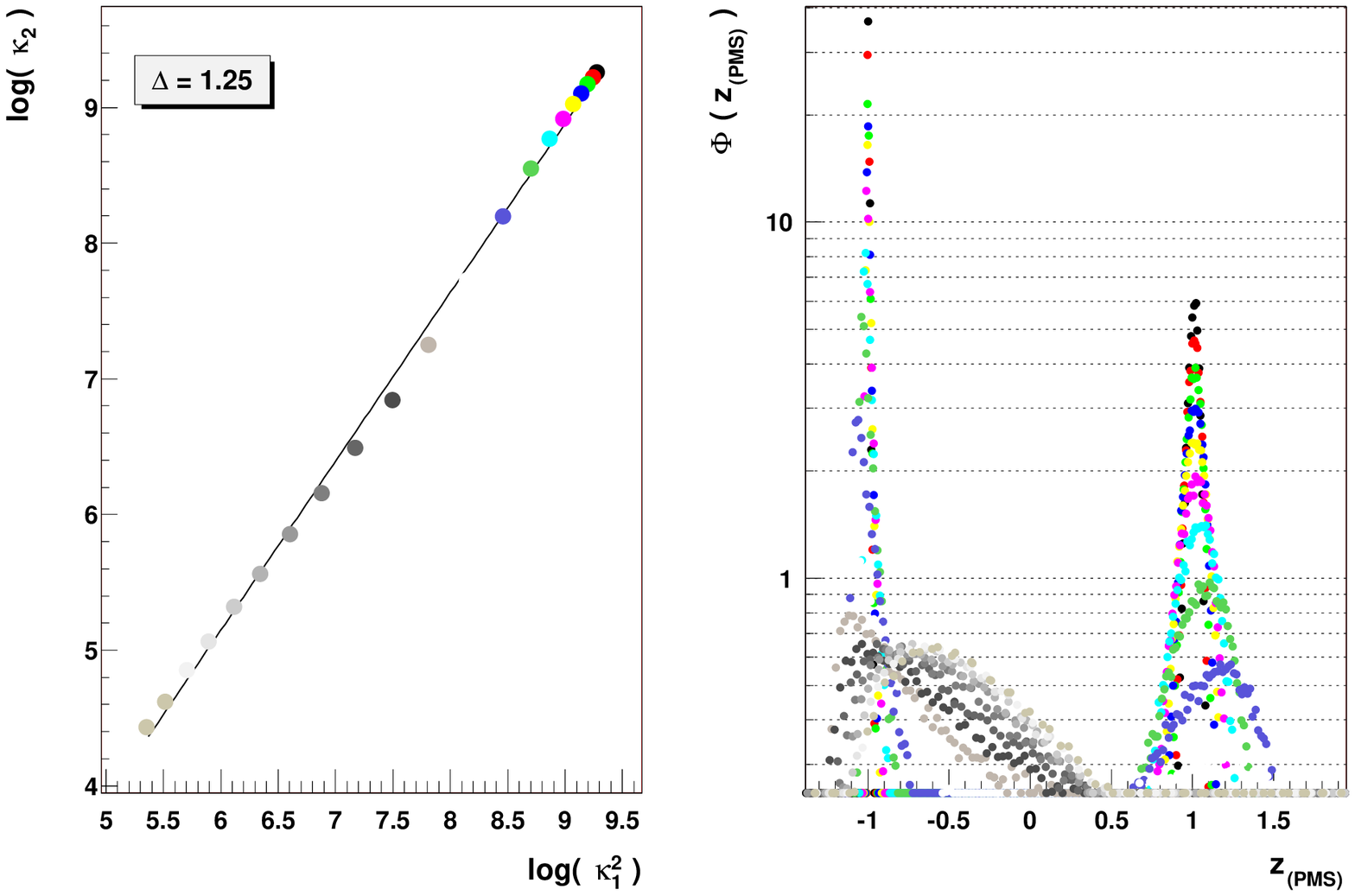,width=5cm}\\
\centering\epsfig{file=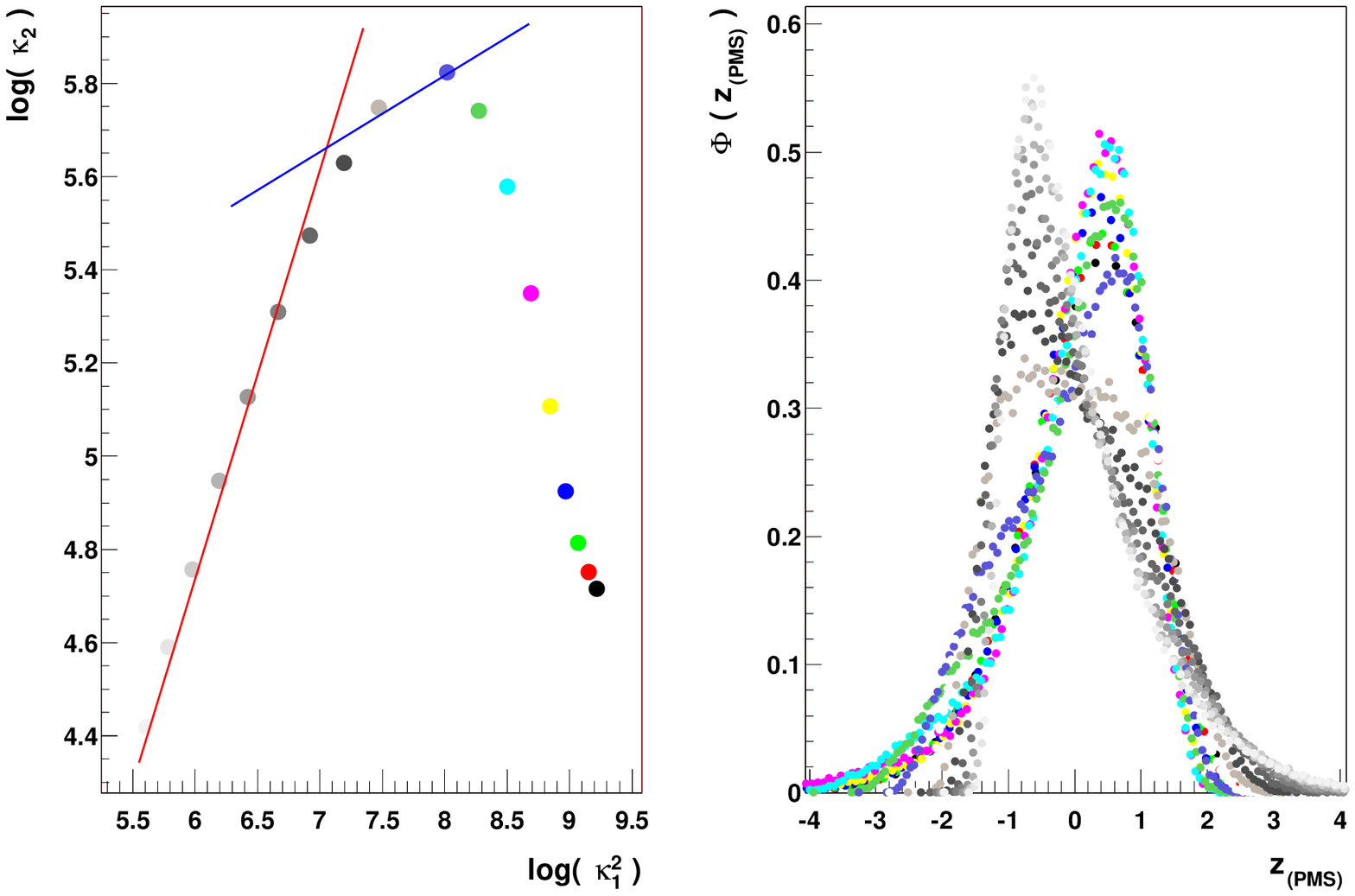,width=5cm}
\caption{Logarithme de la variance de la distribution de taille du plus gros fragment en fonction du logarithme de la valeur moyenne (figure de gauche) et fonctions de scaling (cf. texte) de ces distributions (droite) dans l'ensemble grand-canonique (haut) et canonique (bas).}
\label{scaling} 
\end{figure}

Nous avons appliqué cette méthode d'analyse aux données de notre modèle pour mieux comprendre ce signal. 
 Similairement à la méthode expérimentale, nous avons gardé la taille du réseau constante (6x6x6, ce qui correspond à 108 particules dans l'ensemble canonique).

La figure (\ref{scaling}) montre les résultats obtenus dans les ensembles grand-canonique (haut) et canonique (bas). La partie gauche de chaque sous-figure montre la relation entre le logarithme de la valeur moyenne ($ln(\kappa_{1}^{2})$) et le logarithme de la variance ($ln(\kappa_{2})$) des distributions du plus gros fragment, chaque point correspond à une température (basses températures à droite, hautes à gauche), la partie droite de ces figures superpose les fonctions d'échèlle $\Psi(z_{PMS})$.

Dans l'ensemble grand-canonique, comme nous l'avons vu au paragraphe précédent, les distributions du plus gros fragment dans la zone de coexistence de phase sont bimodales. La variance globale a donc une signification qui n'est plus compatible avec la théorie des fluctuations universelles. Dans cet ensemble les fonctions d'échèlle de ces distributions ne se superposent pas, il n'y a pas d'invariance par changement d'échèlle. La relation entre variance et moyenne est linéaire, aucun accident ne permet dans ce diagrame de déceler la localisation du point critique. Il est plus judicieux par exemple d'utiliser comme nous l'avons vu la disparition de la bimodalité pour localiser le point critique. Seules les distributions correspondant à des températures surcritiques semblent approximativement suivre une loi d'échèlle avec $\Delta \approx 1$.

Dans l'ensemble canonique, les conclusions sont très différentes, la valeur moyenne de la taille du plus gros fragment diminue avec la température, la variance augmente, passe par un maximum puis rediminue 
traduisant l'élargissement de la distribution de plus gros fragments observée au voisinage de la température critique. Les fonctions d'échèlle sont regroupées en deux familles distinctes possédant une asymétrie propre. Les distributions correspondant aux basses températures sont inclinées vers la droite tandis que celles correspondant aux grandes températures sont penchées vers la gauche, au passage de la température critique cette distribution du paramètre d'ordre  s'aplatit avec une forme très large (fluctuation maximale). Assez loin de la température critique, les distributions de chaque famille se ressemblent beaucoup mais ne suivent pas rigoureusement une loi d'échèlle. 
Le $\Delta-scaling$ n'est pas observé bien que le point critique thermodynamique soit exploré par la simulation (point d'intersection des deux droites dans le diagramme de gauche). En effet, même si on accepte d'interpréter la ressemblence des distributions comme une loi d'échèlle déformée par les effets de taille finie, il est clair sur la figure (\ref{scaling}) que la relation entre $ln(A_{big})$ et $ln(\sigma_{A_{big}})$ n'est pas linéaire sauf peut-être pour les températures fortement surcritiques. Toutefois les interpolations linéaires de la figure (\ref{scaling}) montrent que si l'on dispose d'un domaine de variation de température limité, la réduction de la variance due à la loi de conservation sur la masse totale peut être faussement interprétée comme un changement de régime de $\Delta=1$ à $\Delta=\frac{1}{2}$.
%
\section{Conclusion}
Dans cette contribution nous avons montré que les distributions des observables et les moments associés (moyenne, variance ...) dépendent de façon critique de l'ensemble statistique considéré (ie. les contraintes appliquées à notre système). Ceci est spécialement important dans la recherche d'un paramètre d'ordre pour une transition de phase. Une observable peut être paramètre d'ordre pour un ensemble statistique et ne plus l'être pour une autre. En particulier notre étude montre qu'il est difficile de penser que $A_{big}$ puisse être utilisé comme paramètre d'ordre dans les expériences de multifragmentation nucléaire étant donné sa corrélation avec la masse totale du système qui est fortement contrainte dans les expériences. Par contre des observables se basant sur l'asymétrie des distributions du plus gros fragment semblent très prometteuses.
%
%
%
%

\end{document}